\documentclass[12pt]{article}
\usepackage{graphicx,amssymb,epstopdf}

\def\beq{\begin{equation}}
\def\eeq{\end{equation}}
\def\bea{\begin{eqnarray}}
\def\eea{\end{eqnarray}}
\def\nn{\nonumber}
\def\nl{\nonumber\\}
\def\sss{\scriptscriptstyle}

\def\roughly#1{\mathrel{\raise.3ex\hbox
{$#1$\kern-.75em\lower1ex\hbox{$\sim$}}}}
\def\lsim{\roughly<}

\def\sla#1{\raise.15ex\hbox{$/$}\kern-.57em #1}

\def\ks{K_{\sss S}}

\def\bd{B_d^0}

\def\btos{{\bar b} \to {\bar s}}

\def\btopik{B \to \pi K}

\def\ie{{\it i.e.}}

\begin{document}

\begin{titlepage}

\begin{flushright}
UdeM-GPP-TH-09-176 \\
\end{flushright}
\bigskip
\bigskip

\begin{center}
  
{\boldmath\bf\Large Looking for New Physics in $B \to K^* \pi$ and $B \to \rho
  K$ Decays} \\
\bigskip
{\large Cheng-Wei Chiang $^{a,}$\footnote{chengwei@phy.ncu.edu.tw}
and David London $^{b,}$\footnote{london@lps.umontreal.ca}}
\\
\end{center}

\begin{flushleft}
~~~~~~$a$: {\it Department of Physics and Center for
Mathematics and Theoretical Physics,}\\
~~~~~~~~~~{\it National Central University, Chungli,
Taiwan 320, Taiwan;}\\
~~~~~~~~~~{\it Institute of Physics, Academia Sinica,
Taipei, Taiwan 115, Taiwan}\\
~~~~~~$b$: {\it Physique des Particules, Universit\'e de
Montr\'eal,}\\
~~~~~~~~~~{\it C.P. 6128, succ. centre-ville,
Montr\'eal, QC, Canada H3C 3J7}
\end{flushleft}

\date{\today}

\begin{abstract}
  $B \to K^* \pi$ and $B \to \rho K$ decays involve the same quark-level
  processes as $B \to \pi K$.  Analyzing the measurements of the former decays
  might be able to shed additional light on the new-physics hints in the current
  $B \to \pi K$ data.  We perform fits to $B \to K^* \pi$ and $B \to \rho K$
  decays, and find that the data can be accommodated within the standard model.
  However, this agreement is due principally to the large errors in the data,
  particularly the CP-violating asymmetries.  If the errors on the $B \to K^*
  \pi$ and $B \to \rho K$ observables can be reduced, one will have a clearer
  sense of whether new physics is present in these decays.
\end{abstract}


\end{titlepage}


In recent years, experimental data in $\btopik$ decays have shown some
discrepancies with the predictions of the standard model (SM), hinting at
new-physics (NP) contributions in the decay processes \footnote{In the latest
  update of the $\pi K$ puzzle, it was seen that, although NP was hinted at
  $\btopik$ decays, it could be argued that the SM can explain the data, see
  Ref.~\cite{piKupdate}}. This so-called $\pi K$ puzzle has received a great
deal of attention. The NP presumably enters the short-distance pieces of the
decay amplitudes. It should therefore affect other closely-related processes as
well.  In particular, the final-state $VP$ counterparts of $\btopik$ decays ($V$
is a vector meson, $P$ is a pseudoscalar \footnote{There are a great many
  articles on $B\to VP$ decays, though few specifically on $B \to K^* \pi$ and
  $B \to \rho K$.  For general papers on $B\to VP$ decays, see Ref.~\cite{VP1}.
  For analyses more specific to the $B \to K^* \pi$ and $B \to \rho K$ decays,
  see Refs.~\cite{VP2} and \cite{VP3}.}) -- $B \to K^* \pi$ and $B \to \rho K$
decays -- involve essentially the same quark-level processes.  It is thus of
interest to see whether the current experimental data of these decays also
present similar hints of NP. (Even within the SM, these decay modes are useful
in extracting the weak phase $\gamma$ \cite{Gronau,WMS,CWC}.)

The eight $VP$ decay modes can be readily divided into two sets: (I) $B \to K^*
\pi$ decays and (II) $B \to \rho K$ decays, each of which involves a different
set of flavor amplitudes from the other. As is usually done, in our analysis we
neglect the dynamically-suppressed exchange and annihilation amplitudes.  In
addition, the quantum chromodynamics (QCD) penguin amplitude has three pieces,
and can be written as
\begin{eqnarray}
P' & = & V_{ub}^* V_{us} {\tilde P}'_u + V_{cb}^* V_{cs}
{\tilde P}'_c + V_{tb}^* V_{ts} {\tilde P}'_t \nn\\
& = & V_{ub}^* V_{us} ({\tilde P}'_u - {\tilde P}'_c) +
V_{tb}^* V_{ts} ({\tilde P}'_t - {\tilde P}'_c) ~.
\end{eqnarray}
(The primes on the amplitudes indicate $\btos$ transitions.)  In writing the
second line, we have used the unitarity of the Cabibbo-Kobayashi-Maskawa (CKM)
matrix.  In the following, all diagrams (except ${\tilde P}'_c$) are redefined
to absorb the magnitudes of the CKM matrix elements which multiply them, \ie,
$P'_u \equiv |V_{ub}^* V_{us}| {\tilde P}'_u$, $P'_{uc} \equiv |V_{ub}^* V_{us}|
({\tilde P}'_u - {\tilde P}'_c)$, $P'_{tc} \equiv |V_{tb}^* V_{ts}| ({\tilde
  P}'_t - {\tilde P}'_c)$, etc.  Because $|V_{ub}^* V_{us}| \ll |V_{tb}^*
V_{ts}|$, it is expected that $|P'_{uc}| \ll |P'_{tc}|$.  To begin with, we keep
six types of diagrams: the color-allowed tree, denoted by $T'$; the
color-suppressed tree, $C'$; two QCD penguin amplitudes, $P'_{tc}$ and
$P'_{uc}$; the color-allowed EW penguin, $P'_{EW}$; and the color-suppressed EW
penguin, $P_{EW}^{\prime C}$.  We further use a subscript $P$ or $V$ for each
diagram to indicate which final-state meson contains the spectator quark of the
$B$ meson.

Explicitly, we have for the decay modes in Set (I):
\begin{eqnarray}
\label{eq:I}
A_I^{0+} &=& - P'_{tc,P} + P'_{uc,P} e^{i\gamma} + \frac13
P_{EW,P}^{\prime C} ~, \nl
\sqrt{2} A_I^{+0} &=& P'_{tc,P} - P'_{uc,P} e^{i\gamma} +
P'_{EW,V} - T'_P e^{i\gamma} - C'_V e^{i\gamma} + \frac23
P_{EW,P}^{\prime C} ~, \nl
A_I^{+-} &=& P'_{tc,P} - P'_{uc,P} e^{i\gamma} - T'_P
e^{i\gamma} + \frac23 P_{EW,P}^{\prime C} ~, \nl
\sqrt{2} A_I^{00} &=& - P'_{tc,P} + P'_{uc,P} e^{i\gamma} +
P'_{EW,V} - C'_V e^{i\gamma} + \frac13 P_{EW,P}^{\prime C} ~;
\end{eqnarray}
and for the decay modes in Set (II):
\begin{eqnarray}
\label{eq:II}
\sqrt{2} A_{II}^{0+} &=& P'_{tc,V} - P'_{uc,V} e^{i\gamma} +
P'_{EW,P} - T'_V e^{i\gamma} - C'_P e^{i\gamma} + \frac23
P_{EW,V}^{\prime C} ~, \nl
A_{II}^{+0} &=& - P'_{tc,V} + P'_{uc,V} e^{i\gamma} + \frac13
P_{EW,V}^{\prime C} ~, \nl
A_{II}^{-+} &=& P'_{tc,V} - P'_{uc,V} e^{i\gamma} - T'_V
e^{i\gamma} + \frac23 P_{EW,V}^{\prime C} ~, \nl
\sqrt{2} A_{II}^{00} &=& - P'_{tc,V} + P'_{uc,V} e^{i\gamma}
+ P'_{EW,P} - C'_P e^{i\gamma} + \frac13 P_{EW,V}^{\prime C}
~;
\end{eqnarray}
In the amplitudes $A_\alpha^{ij}$ ($\alpha \in \left\{ I, II \right\}$ and $i,j
\in \left\{ 0,+ \right\}$) appearing in Eqs.~(\ref{eq:I}) and (\ref{eq:II}), the
subscript $\alpha$ denotes the decay set and the superscripts $ij$ denote the
electric charges of the vector and pseudoscalar mesons, respectively.  Here we
have explicitly written the weak phase associated with each amplitude (including
the minus sign of $V_{tb}^* V_{ts}$ in the QCD and EW penguins), while the
diagrams contain the strong phases.

In fact, there are relations among the various diagrams.  The ratios of Wilson
coefficients $c_{10}/c_9$ and $c_2/c_1$ are equal, to a good approximation (the
difference is only about 3\%).  In the limit in which this equality is exact,
the relations read \cite{Gronau}
\bea
P'_{EW,M} & = & -\frac32 {c_9 \over c_1} \left( T'_M +
P'_{u,M'} \right) ~~~~ M' \ne M ~, \nl
P_{EW,M}^{\prime C} & = & -\frac32 {c_9 \over c_1} \left(
C'_M - P'_{u,M} \right) ~.
\eea
Now, the diagram $P'_u$ does not appear in the $B$-decay amplitudes.  Thus, in
order to use these relations, an approximation must be made.  We consider two
possibilities.  In the first, we neglect all factors of $P'_u$ and $P'_{uc}$.
In the second, we replace the $P'_u$ in the above relations by $P'_{uc}$.  Both
approximations are justified because $|P'_{u}|$ and $|P'_{uc}|$ are expected to
be small.  In either case, the relations relate some diagrams appearing in $B
\to K^* \pi$ decays to some in $B \to \rho K$ decays.  That is, the fits must be
performed to measurements of both decays simultaneously.  We call the fits
corresponding to the two approximations Fit1 and Fit2.  Also, throughout this
paper, the strong phases of the amplitudes are defined with respect to $T'_P$.

Current experimental measurements of the observables in $B \to K^* \pi$ and $B
\to \rho K$ decays are given in Table~\ref{tab:data}. In the first
approximation, where $P'_{uc}$ is neglected, the direct CP asymmetries
($A_{CP}$'s) of $K^{*0} \pi^+$ and $\rho^0 K^+$ are identically zero within this
framework.  Therefore, it is of no use including them in the $\chi^2$ functions
as they impose no constraint on the theory parameters.  We are then left with 15
and 17 observables in Fit1 and Fit2, respectively.

\begin{table}[t]
\begin{center}
\begin{tabular}{lccccc}
\hline\hline
Set & Mode & BR ($\times 10^{-6}$) & $A_{CP}$ & $S_{CP}$ \\ 
\hline
(I) & $B^+ \to K^{*0} \pi^+$
    & $10.0 \pm 0.8$ & $-0.020^{+0.067}_{-0.061}$ & -- \\
    & $B^+ \to K^{*+} \pi^0$ 
    & $6.9 \pm 2.3$ & $0.04 \pm 0.29$ & -- \\
    & $B^0 \to K^{*+} \pi^-$
    & $10.3 \pm 1.1$ & $-0.25 \pm 0.11$ & -- \\
    & $B^0 \to K^{*0} \pi^0$
    & $2.4 \pm 0.7$ & $-0.15 \pm 0.12$ & -- \\
\hline
(II) & $B^+ \to \rho^0 K^+$
    & $3.81^{+0.48}_{-0.46}$ & $0.419^{+0.081}_{-0.104}$ & -- \\
    & $B^+ \to \rho^+ K^0$
    & $8.0^{+1.5}_{-1.4}$ & $-0.12 \pm 0.17$ & -- \\
    & $B^0 \to \rho^- K^+$
    & $8.6^{+0.9}_{-1.1}$ & $0.15 \pm 0.06$ & -- \\
    & $B^0 \to \rho^0 K^0$
    & $5.4^{+0.9}_{-1.0}$ & $0.01 \pm 0.20$ & $0.63^{+0.17}_{-0.21}$ \\
\hline\hline
\end{tabular}
\end{center}
\caption{Branching ratios, direct CP asymmetries $A_{CP}$,
  and mixing-induced CP asymmetry $S_{CP}$ (if applicable and
  measured) in $B \to K^* \pi$ and $B \to \rho K$ decays.
  Quoted values are taken from Ref.~\cite{HFAG} that
  weight-averages individual measurements \cite{EXP1,EXP2,EXP3,EXP4,EXP5,EXP6,%
    EXP7,EXP8,EXP9,EXP10,EXP11,EXP12,EXP13,EXP14}.}
\label{tab:data}
\end{table}

In both Fit1 and Fit2, one of the theoretical parameters is the weak phase
$\gamma$.  We have two options in treating it.  One possibility is to extract
$\gamma$ from the $B \to K^* \pi$ and $B \to \rho K$ measurements.  This value
can then be compared with that obtained from independent measurements
\cite{CKMfitter}:
\beq
\gamma = (66.8^{+5.4}_{-3.8})^\circ ~.
\label{gammaval}
\eeq
If the two values of $\gamma$ disagree, this will be a hint of NP.
Alternatively, one can impose the value of $\gamma$ of Eq.~(\ref{gammaval}) by
adding a constraint to the fits.  We perform fits using both of these options.
Fit1-a and Fit2-a extract $\gamma$ from the data; Fit1-b and Fit2-b have the
additional $\gamma$ constraint (in which the asymmetrical errors are averaged).

In all fits, there will be a hint of NP if the extracted values of the diagrams
disagree with the SM calculations.  In the QCD factorization (QCDF) approach,
the default parameters have the following ratios of amplitudes
\cite{Beneke:2003zv}:
\begin{eqnarray}
\label{eq:QCDF}
|C'_V / T'_P| = 0.16 \pm 0.11 ~, \mbox{ and }
|T'_P / P'_{tc,P}| = 0.59 \pm 0.12 ~; \nl
|C'_P / T'_V| = 0.20 \pm 0.13 ~, \mbox{ and }
|T'_V / P'_{tc,V}| = 0.55 \pm 0.26 ~.
\end{eqnarray}

We begin with Fit1-a.  Here, the diagram $P'_{uc}$ is neglected.  There are
therefore a total of 11 hadronic theoretical parameters: 6 amplitude magnitudes,
and 5 relative strong phases.  In addition, we have the unknown weak phase
$\gamma$.  Finally, the mixing-induced CP asymmetry $S_{CP}(\rho^0 K^0)$ depends
also on the weak phase $\beta$.  For this, we impose the additional constraint
$\beta = (21.66^{+0.95}_{-0.87})^\circ$ \cite{CKMfitter}, which is obtained
mainly from the measurement of CP violation in $\bd(t)\to J/\Psi \ks$ and other
${\bar b} \to {\bar c} c {\bar s}$ decays.  We average the asymmetrical errors
on $\beta$. The degree of freedom ($d.o.f.$) is therefore 3 in this fit.

Before presenting the results of this fit, we note the following.  There are
quite a few solutions, many of which suggest the presence of new physics (NP).
While we cannot be absolutely sure that one of these is not the true solution,
we make the assumption that the NP, if present, is not enormous.  If it were, it
probably would already have been seen elsewhere.  The large-NP scenarios can be
eliminated by demanding that an acceptable solution respect certain constraints
on the ratios of magnitudes of diagrams.  Now, as we will see, the errors in the
fits of the theoretical parameters are quite large.  If we consider the full
1$\sigma$ range of these parameters, all potential solutions will respect the
constraints.  For this reason, we concentrate only on the central values.  In
particular, we require that the central values of any solution do not violate
the ratios of Eq.~(\ref{eq:QCDF}) by a large amount.  We impose the conservative
constraints
\beq
|C'_V / T'_P|, ~|C'_P / T'_V| \le 1 ~, \mbox{ and } |T'_P /
 P'_{tc,P}|, ~|T'_V / P'_{tc,V}| \le 2 ~.
\label{NPnothuge}
\eeq
If any one of these is violated, the solution is excluded.  There are a great
many possible solutions.  But if we restrict $\chi^2_{min}$ to be reasonably
small, there are only 14.  The above constraints eliminate 12 of these.

\begin{table}[p]
\center
\begin{tabular}{ccccccccc}
\hline\hline
$\chi^2_{min}$ & $|P'_{tc,P}|$ & $|P'_{tc,V}|$ & $|T'_P|$ &
$|T'_V|$ \\
\hline
0.98 & $32.9^{+1.5}_{-1.6}$ & $29.0^{+3.2}_{-2.5}$ &
  $13.3^{+14.6}_{-7.6}$ & $11.7^{+4.9}_{-5.5}$ \\
\hline
1.08 & $32.4 \pm 1.6$ & $29.3^{+3.7}_{-2.6}$ &
$13.6^{+14.2}_{-7.7}$ & $12.8^{+8.0}_{-5.4}$ \\
\hline\hline
$|C'_P|$ & $|C'_V|$ & $\delta_{P'_{tc,P}}$ &
$\delta_{P'_{tc,V}}$ & $\delta_{T'_V}$ \\
\hline
$1.6^{+4.0}_{-1.6}$ & $2.2^{+9.7}_{-1.6}$ &
$(-18^{+12}_{-38})^\circ$ & $(128^{+43}_{-22})^\circ$ &
$(-36^{+40}_{-21})^\circ$ \\
\hline
$0.9^{+4.0}_{-0.9}$ & $2.9^{+8.8}_{-2.3}$ &
$(-161^{+35}_{-12})^\circ$ & $(-121^{+18}_{-39})^\circ$ &
$(-136^{+17}_{-46})^\circ$ \\
\hline\hline
& $\delta_{C'_P}$ & $\delta_{C'_V}$ & $\beta$ & $\gamma$ \\
\hline
& $(-27 \pm 180)^\circ$ & $(-119^{+96}_{-75})^\circ$ &
$(21.67^{+1.0}_{-1.0})^\circ$ &
$(82.9^{+32.2}_{-14.7})^\circ$ \\
\hline
& $(76 \pm 180)^\circ$ & $(144^{+51}_{-118})^\circ$ &
$(21.65 \pm 0.91)^\circ$ &
$(91.5^{+19.3}_{-32.8})^\circ$ \\
\hline\hline
\end{tabular}
\caption{Solutions of Fit1-a ($d.o.f. = 3$) to the decay
observables in $B \to K^* \pi$ and $B \to \rho K$ decays.
The amplitude magnitudes are quoted in units of eV.}
\label{tab:1-a}
\vspace{1cm}
\begin{tabular}{ccccccccc}
\hline\hline
$\chi^2_{min}$ & $|P'_{tc,P}|$ & $|P'_{tc,V}|$ & $|T'_P|$ &
$|T'_V|$ \\
\hline
1.66 & $32.0^{+1.3}_{-1.5}$ & $31.7^{+1.5}_{-1.7}$ &
  $8.6^{+4.4}_{-3.6}$ & $16.0^{+6.1}_{-5.3}$ \\
\hline
1.99 & $32.9^{+2.0}_{-1.4}$ & $27.9^{+2.0}_{-2.2}$ &
  $27.5^{+6.1}_{-16.9}$ & $6.9^{+4.3}_{-3.8}$ \\
\hline
2.06 & $33.9^{+1.1}_{-2.4}$ & $27.7^{+2.3}_{-2.6}$ &
  $15.6^{+14.9}_{-5.2}$ & $11.7^{+6.1}_{-8.7}$ \\
\hline\hline
$|C'_P|$ & $|C'_V|$ & $\delta_{P'_{tc,P}}$ &
$\delta_{P'_{tc,V}}$ & $\delta_{T'_V}$ \\
\hline
$2.5^{+3.3}_{-2.4}$ & $6.4^{+5.2}_{-4.2}$ &
$(-143^{+29}_{-16})^\circ$ & $(-138^{+27}_{-36})^\circ$ &
$(-149^{+26}_{-32})^\circ$ \\
\hline
$2.0^{+5.4}_{-2.0}$ & $10.9^{+8.6}_{-10.0}$ &
$(-8^{+6}_{-20})^\circ$ & $(109^{+49}_{-12})^\circ$ &
$(-36^{+41}_{-17})^\circ$ \\
\hline
$4.6^{+2.8}_{-4.6}$ & $3.2^{+7.0}_{-2.4}$ &
$(-15 \pm 13)^\circ$ & $(141^{+17}_{-45})^\circ$ &
$(-21^{+20}_{-32})^\circ$ \\
\hline\hline
& $\delta_{C'_P}$ & $\delta_{C'_V}$ & $\beta$ & $\gamma$ \\
\hline
& $(79^{+107}_{-51})^\circ$ & $(-163^{+35}_{-29})^\circ$ &
$(21.67 \pm 0.91)^\circ$ & $(67.2 \pm 4.6)^\circ$
\\
\hline
& $(-57 \pm 180)^\circ$ & $(-170^{+148}_{-12})^\circ$ &
$(21.67 \pm 0.91)^\circ$ & $(67.3^{+5.3}_{-4.6})^\circ$
\\
\hline
& $(-22 \pm 180)^\circ$ & $(-63^{+43}_{-118})^\circ$ &
$(21.75 \pm 0.91)^\circ$ & $(68.4 \pm 4.3)^\circ$
\\
\hline\hline
\end{tabular}
\caption{Solutions of Fit1-b ($d.o.f. = 4$) to the decay
observables in $B \to K^* \pi$ and $B \to \rho K$ decays.
The amplitude magnitudes are quoted in units of eV.}
\label{tab:1-b}
\end{table}

As we will see, one of the conclusions of our analysis is that it is important
to reduce the experimental errors on the measurements.  When this is done, the
$\chi^2$ distribution will be deeper, and there will be far fewer minima.

The two solutions which are retained are shown in Table~\ref{tab:1-a}. Both of
them have very good fit qualities -- $81\%$ and $78\%$, respectively. (The
quality of fit depends on $\chi^2_{min}$ and $d.o.f.$ individually. A value of
50\% or more is a good fit; fits which are substantially less than 50\% are
poorer.)

We now turn to Fit1-b.  In this fit, the additional constraint on $\gamma$
[Eq.~(\ref{gammaval})] is imposed, increasing the $d.o.f.$ by one to 4.  The
assumption that any NP is not enormous implies again that the conditions of
Eq.~(\ref{NPnothuge}) are required, leading to the elimination of 10 of 13
possible solutions. The three that are retained are given in
Table~\ref{tab:1-b}. They have good fit qualities -- $64\%$, $58\%$ and $56\%$,
respectively.

In Fit2-a, we keep the diagrams $P'_{uc,P}$ and $P'_{uc,V}$.  Their addition
increases the number of hadronic theoretical parameters by 4 (2 magnitudes, 2
strong phases).  However, the direct CP asymmetries in $B \to K^{*0} \pi^+$ and
$B \to \rho^0 K^+$ are no longer zero, leading to an increase in the number of
observables by 2.  The net effect is that the $d.o.f.$ is reduced by 2 with
respect to that of Fit1-a, \ie, it is 1. In the SM, it is expected that
$|P'_{uc,P} /
P'_{tc,P}|, |P'_{uc,V} / P'_{tc,V}| \sim 5\%$ (in fact, it has been argued in
Ref.~\cite{Kim} that the $|P'_{uc} / P'_{tc}|$ ratio is smaller). In order to
avoid too-large NP, in addition to the constraints of Eq.~(\ref{NPnothuge}), we
also require that
\beq
|P'_{uc,P} / P'_{tc,P}|, ~|P'_{uc,V} / P'_{tc,V}| < 1 ~.
\label{PucNPnothuge}
\eeq
The imposition of all these conditions leads to the elimination of 17 of 20
possible solutions.  The three that are kept are given in
Table~\ref{tab:2-a}. The qualities of fit are good -- $64\%$, $62\%$ and $61\%$,
respectively.

\begin{table}[p]
\center
\begin{tabular}{cccccccc}
\hline\hline
$\chi^2_{min}$ & $|P'_{tc,P}|$ & $|P'_{tc,V}|$ & $|T'_P|$ &
$|T'_V|$ \\
\hline
0.22 & $31.1^{+25.6}_{-9.1}$ & $29.9^{+15.6}_{-8.1}$ &
  $28.5^{+24.2}_{-16.3}$ & $11.8^{+9.5}_{-7.9}$ \\
\hline
0.24 & $33.9^{+12.9}_{-24.6}$ & $24.5^{+7.4}_{-28.1}$ &
  $21.6^{+18.0}_{-16.4}$ & $15.1^{+34.4}_{-13.1}$ \\
\hline
0.26 & $33.6^{+7.6}_{-6.6}$ & $22.0^{+13.4}_{-12.5}$ &
  $28.7^{+23.5}_{-16.7}$ & $11.6^{+16.8}_{-8.1}$ \\
\hline\hline
$|C'_P|$ & $|C'_V|$ & $|P'_{uc,P}|$ & $|P'_{uc,V}|$ &
$\delta_{P'_{tc,P}}$ \\
\hline
$3.8^{+25.3}_{-3.8}$ & $20.4^{+26.9}_{-20.4}$ & $10.7^{+22.6}_{-10.7}$ &
$3.3^{+19.1}_{-3.3}$ & $(-170 \pm 16)^\circ$ \\
\hline
$14.1^{+20.1}_{-14.1}$ & $5.3^{+35.3}_{-5.3}$ & $15.7^{+38.8}_{-30.5}$ &
$12.6^{+29.4}_{-12.6}$ & $(13^{+34}_{-50})^\circ$ \\
\hline
$8.4^{+17.0}_{-8.4}$ & $6.4^{+30.9}_{-6.4}$ & $0.8^{+17.4}_{-0.8}$ &
$10.1^{+12.6}_{-10.1}$ & $(-176 \pm 22)^\circ$ \\
\hline\hline
$\delta_{P'_{tc,V}}$ & $\delta_{T'_V}$ & $\delta_{C'_P}$ & $\delta_{C'_V}$ &
$|\delta_{uc,P}|$ \\
\hline
$(-106^{+30}_{-61})^\circ$ & $(-151^{+59}_{-117})^\circ$ & $(-33 \pm 180)^\circ$
& $(179 \pm 180)^\circ$ & $(-169 \pm 180)^\circ$ \\
\hline
$(114^{+55}_{-23})^\circ$ & $(96^{+199}_{-160})^\circ$ & $(-69 \pm 180)^\circ$ &
$(116 \pm 180)^\circ$ & $(20^{+38}_{-50})^\circ$ \\
\hline
$(-156^{+66}_{-19})^\circ$ & $(146^{+90}_{-80})^\circ$ & $(-128 \pm 180)^\circ$
& $(175 \pm 180)^\circ$ & $(-157 \pm 180)^\circ$ \\
\hline\hline
$|\delta_{uc,V}|$ & $\beta$ & $\gamma$ \\
\hline
$(-59 \pm 180)^\circ$ & $(21.66 \pm 0.91)^\circ$ &
$(73.8^{+70.5}_{-47.3})^\circ$ \\
\hline
$(-75 \pm 180)^\circ$ & $(21.67 \pm 0.91)^\circ$ &
$(62.1^{+62.5}_{-40.2})^\circ$ \\
\hline
$(-136 \pm 180)^\circ$ & $(21.67 \pm 0.92)^\circ$ &
$(116.2^{+25.6}_{-92.0})^\circ$ \\
\hline\hline
\end{tabular}
\caption{Solutions of Fit2-a ($d.o.f. = 1$) to the decay
observables in $B \to K^* \pi$ and $B \to \rho K$ decays.
The amplitude magnitudes are quoted in units of eV.}
\label{tab:2-a}
\vspace{1cm}
\begin{tabular}{cccccccc}
\hline\hline
$\chi^2_{min}$ & $|P'_{tc,P}|$ & $|P'_{tc,V}|$ & $|T'_P|$ &
$|T'_V|$ \\
\hline
0.26 & $32.8^{+3.0}_{-15.4}$ & $25.3^{+5.7}_{-19.7}$ &
  $20.1^{+10.0}_{-12.6}$ & $13.2^{+23.6}_{-11.1}$ \\
\hline
0.81 & $32.6^{+3.9}_{-5.8}$ & $29.1^{+2.6}_{-7.8}$ &
  $26.5^{+18.2}_{-20.9}$ & $16.7^{+14.2}_{-15.3}$ \\
\hline\hline
$|C'_P|$ & $|C'_V|$ & $|P'_{uc,P}|$ & $|P'_{uc,V}|$ &
$\delta_{P'_{tc,P}}$ \\
\hline
$13.1^{+21.1}_{-13.1}$ & $3.5^{+21.8}_{-3.5}$ & $14.1^{+19.0}_{-14.1}$ &
$11.6^{+22.8}_{-11.6}$ & $(10 \pm 33)^\circ$ \\
\hline
$4.2^{+15.7}_{-4.2}$ & $14.2^{+17.7}_{-14.2}$ & $8.4^{+29.2}_{-8.4}$ &
$6.3^{+29.0}_{-6.3}$ & $(-15^{+32}_{-31})^\circ$ \\
\hline\hline
$\delta_{P'_{tc,V}}$ & $\delta_{T'_V}$ & $\delta_{C'_P}$ & $\delta_{C'_V}$ &
$|\delta_{uc,P}|$ \\
\hline
$(112^{+47}_{-18})^\circ$ & $(93^{+203}_{-154})^\circ$ & $(-71 \pm 180)^\circ$ &
$(105 \pm 180)^\circ$ & $(17 \pm 180)^\circ$ \\
\hline
$(108^{+50}_{-14})^\circ$ & $(-50^{+161}_{-15})^\circ$ & $(112 \pm 180)^\circ$ &
$(-159 \pm 180)^\circ$ & $(-14 \pm 180)^\circ$ \\
\hline\hline
$|\delta_{uc,V}|$ & $\beta$ & $\gamma$ \\
\hline
$(-77 \pm 180)^\circ$ & $(21.67 \pm 0.91)^\circ$ & $(66.7^{+4.6}_{-4.6})^\circ$ \\
\hline
$(115 \pm 180)^\circ$ & $(21.68 \pm 0.92)^\circ$ & $(66.9^{+4.6}_{-4.6})^\circ$ \\
\hline\hline
\end{tabular}
\caption{Solutions of Fit2-b ($d.o.f. = 2$) to the decay
observables in $B \to K^* \pi$ and $B \to \rho K$ decays.
The amplitude magnitudes are quoted in units of eV.}
\label{tab:2-b}
\end{table}

Finally, we have Fit2-b.  This is just like Fit2-a, except that the value of
$\gamma$ is fixed by Eq.~(\ref{gammaval}).  This increases the $d.o.f.$ by one
to 2.  The constraints of Eqs.~(\ref{NPnothuge}) and (\ref{PucNPnothuge})
eliminate 8 of 10 possible solutions. The two that are retained are given in
Table~\ref{tab:2-b}. The fit qualities are $88\%$ and $67\%$, respectively.

We can now examine the various fit solutions of
Tables~\ref{tab:1-a}-\ref{tab:2-b}.  We separate them into three categories,
depending on the level of disagreement with independent measurements
[Eq.~(\ref{gammaval})] or the predictions of the SM [Eq.~(\ref{eq:QCDF})].
There are SM fits (discrepancies at the level of $\lsim 2\sigma$), marginal fits
(discrepancies at the level of $\sim 3\sigma$), and fits which indicate NP
(discrepancies at the level of $\sim 5\sigma$).  If we just concentrate on the
central values of the theoretical parameters found in the fits, we have all
three types, shown in Table~\ref{typeoffit}.  Of the ten solutions in the
Tables, only two are in agreement with the SM.  In fact, six show clear signs of
NP.  If Table~\ref{typeoffit} represented the actual situation, we would
conclude that $B \to K^* \pi$ and $B \to \rho K$ decays might be showing signs
of new physics, just like $B \to \pi K$.

\begin{table}[thb]
\center
\begin{tabular}{ccccccccc}
\hline\hline

Type of fit & Table & $\chi^2_{min}$ & Problem \\
\hline
SM & \ref{tab:1-b} & 1.99 \\
   & \ref{tab:1-b} & 2.06 \\
\hline
marginal & \ref{tab:1-a} & 0.98 & $\gamma$ \\
         & \ref{tab:2-b} & 0.81 & $|C'_V/T'_P|$ \\
\hline
NP & \ref{tab:1-a} & 1.08 & $\gamma$ \\
   & \ref{tab:1-b} & 1.66 & $|C'_V/T'_P|$ \\
   & \ref{tab:2-a} & 0.22 & $|C'_V/T'_P|$ \\
   & \ref{tab:2-a} & 0.24 & $|C'_P/T'_V|$ \\
   & \ref{tab:2-a} & 0.26 & $\gamma$ \\
   & \ref{tab:2-b} & 0.26 & $|C'_P/T'_V|$ \\
\hline\hline
\end{tabular}
\caption{Type of fits in Tables~\ref{tab:1-a}-\ref{tab:2-b},
using only the central values of the theoretical parameters.}
\label{typeoffit}
\end{table}

Unfortunately, this is not the case.  In the fits, the theoretical parameters
have errors associated with them, shown in Tables~\ref{tab:1-a}-\ref{tab:2-b}.
The errors are so large that, when they are taken into account, all solutions
agree with the SM.  This can be seen clearly in the following figures.

For Fit1-a, Fig.~\ref{fig-Fit1-a} shows the correlation between $\gamma$ and
$|C'_P / T'_V|$ (left plot) or $|T'_P / P'_P|$ (right plot).  In both plots, the
shaded region is given by the 1$\sigma$ ranges of Eqs.~(\ref{gammaval}) and
(\ref{eq:QCDF}).  For the central values of the solutions in Table~\ref{tab:1-a}
((red) dots), we see that $\gamma$ is only marginally consistent with one of
them, and inconsistent with the other, indicating the presence of NP.  This is
what Table~\ref{typeoffit} also states.  (There are no inconsistencies with the
ratios of diagrams.) However, these plots also show the allowed regions when the
fit errors on the theoretical parameters are taken into account: the contours of
$\Delta\chi^2 = 1$ and $2$ from the best solution ($\chi^2_{min} = 0.98$) are
indicated by the solid (blue) and dashed (green) curves.  We see quite clearly
that the shaded area is mostly contained in the ``solid region,'' and is
entirely in the ``dashed region.'' That is, when we include the fit errors,
there is no inconsistency -- both solutions are in agreement with the SM.

\begin{figure}[p]
\centering{%
\includegraphics[width=0.45\textwidth]{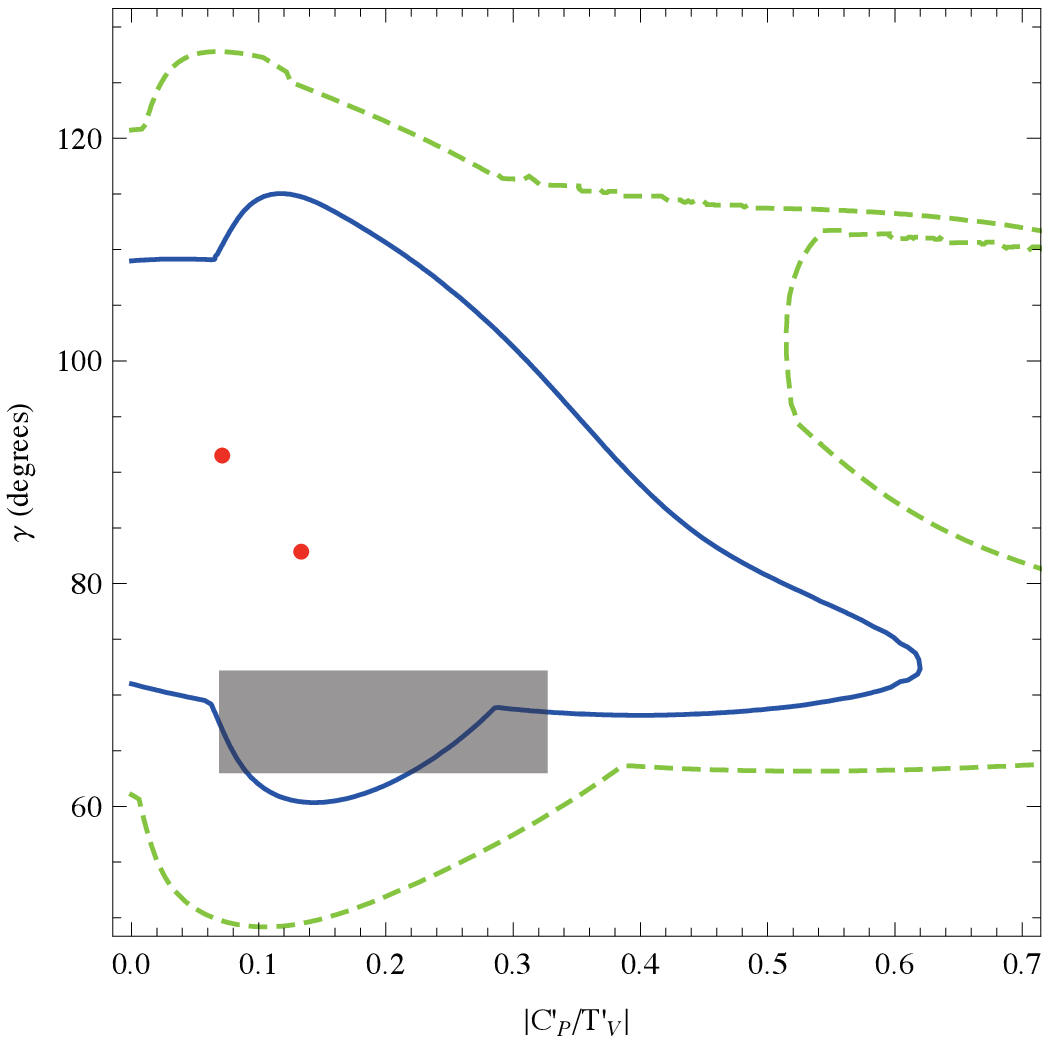}
\hspace{0.5cm}
\includegraphics[width=0.45\textwidth]{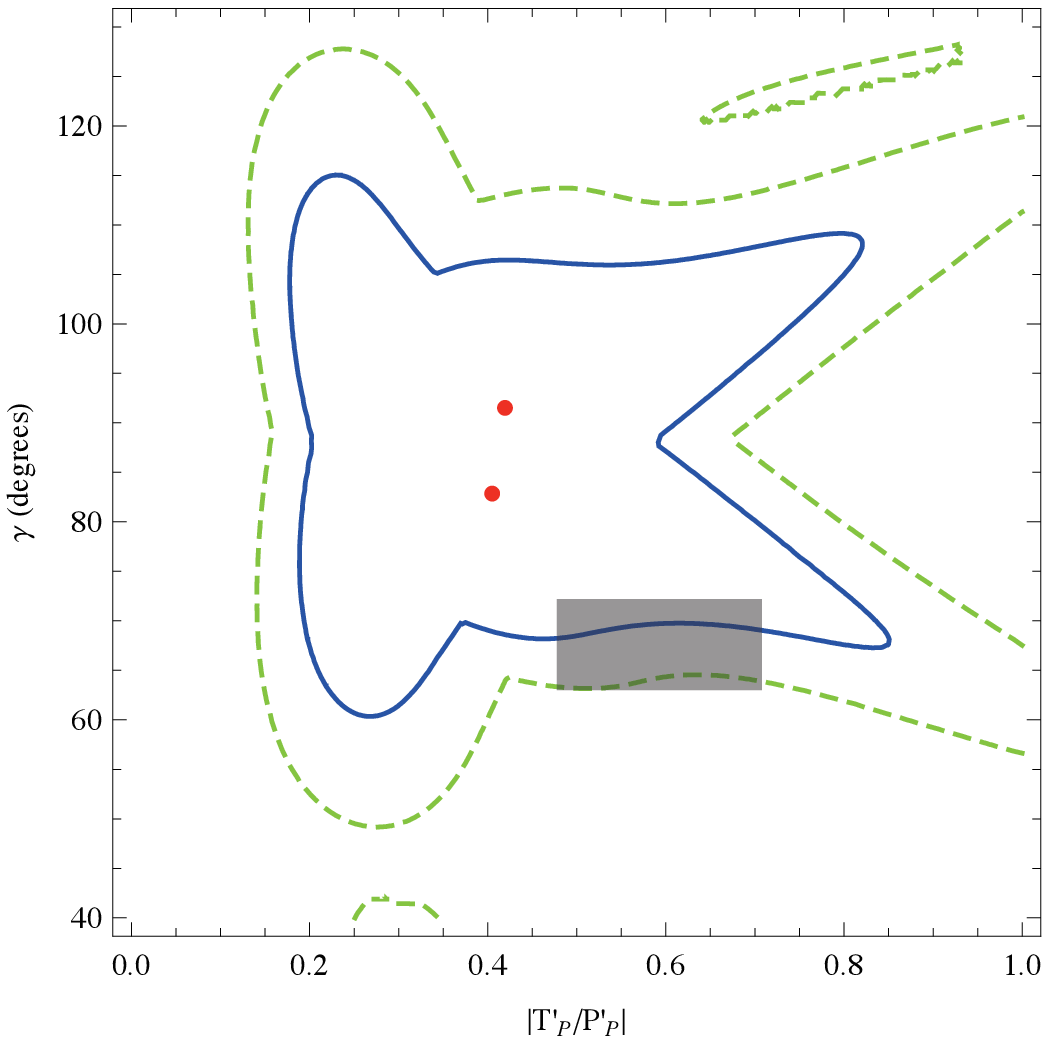}}
\caption{Fit1-a: correlation between the weak phase $\gamma$
  and the ratio $|C'_P / T'_V|$ (left plot) or $|T'_P /
  P'_P|$ (right plot). The (red) dots correspond to the
  solutions listed in Table~\ref{tab:1-a}.  The contours of
  $\Delta\chi^2 = 1$ and $2$ from the best solution
  ($\chi^2_{min} = 0.98$) are indicated by the solid (blue)
  and dashed (green) curves.  The shaded region is given by
  the 1$\sigma$ ranges of Eqs.~(\ref{gammaval}) and
  (\ref{eq:QCDF}) for comparison.}
\label{fig-Fit1-a}
\vspace{1cm}
\centering{%
\includegraphics[width=0.45\textwidth]{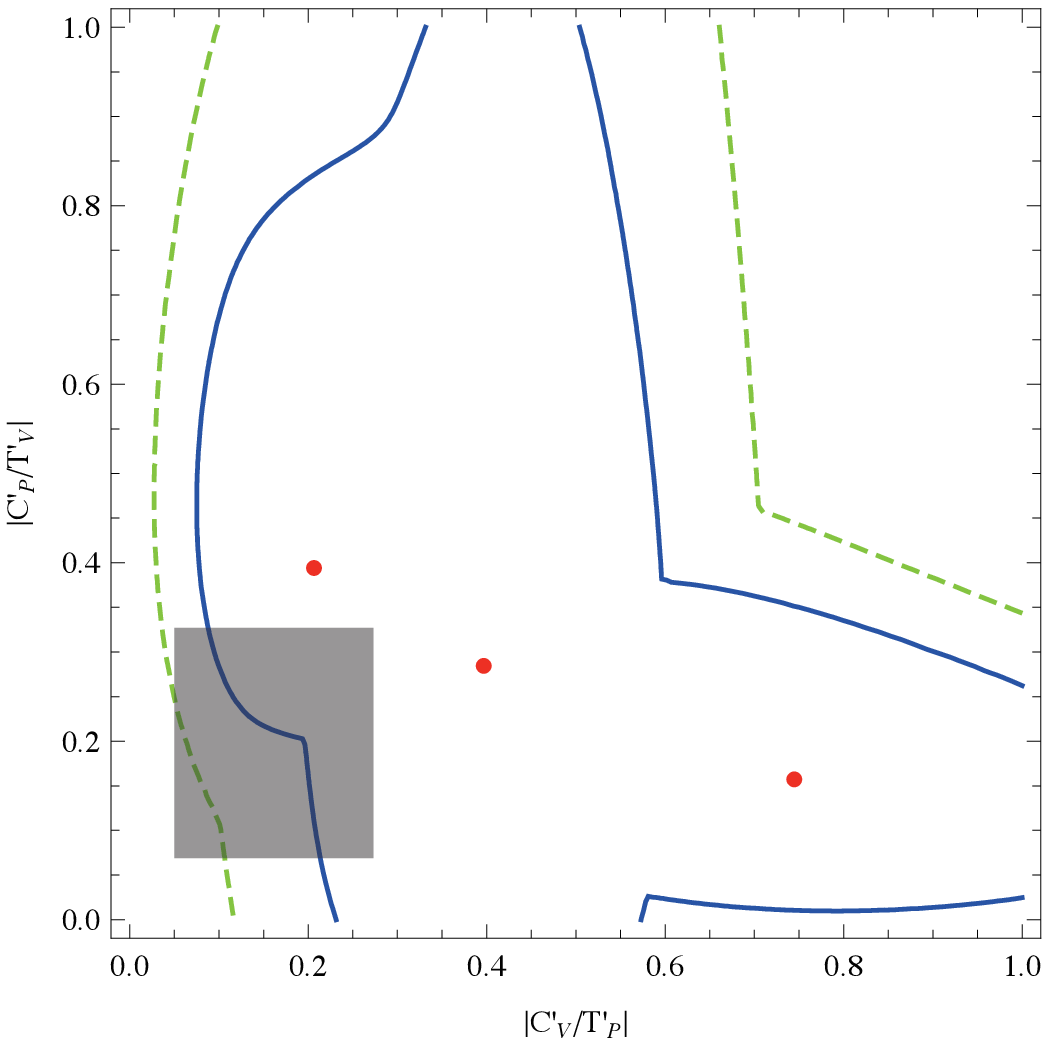}}
\caption{Fit1-b: correlation between the ratios $|C'_P /
  T'_V|$ and $|C'_V / T'_P|$. The (red) dots correspond to
  the solutions listed in Table~\ref{tab:1-b}.  The contours
  of $\Delta\chi^2 = 1$ and $2$ from the best solution
  ($\chi^2_{min} = 1.66$) are indicated by the solid (blue)
  and dashed (green) curves.  The shaded region is given by
  the 1$\sigma$ ranges of Eqs.~(\ref{gammaval}) and
  (\ref{eq:QCDF}) for comparison.}
\label{fig-Fit1-b}
\end{figure}

The situation is similar in Fig.~\ref{fig-Fit1-b}.  Here we see the correlation
between the ratios $|C'_P / T'_V|$ and $|C'_V / T'_P|$ for Fit1-b.  As indicated
in Table~\ref{typeoffit}, the shaded region is consistent with two central
values, and inconsistent with the third. However, when fit errors are included,
the shaded area is included in the allowed regions (essentially completely for
the ``dashed region'').  Once again, when the fit errors are taken into account,
all solutions are in agreement with the SM.

\begin{figure}[thb]
\centering{%
\includegraphics[width=0.45\textwidth]{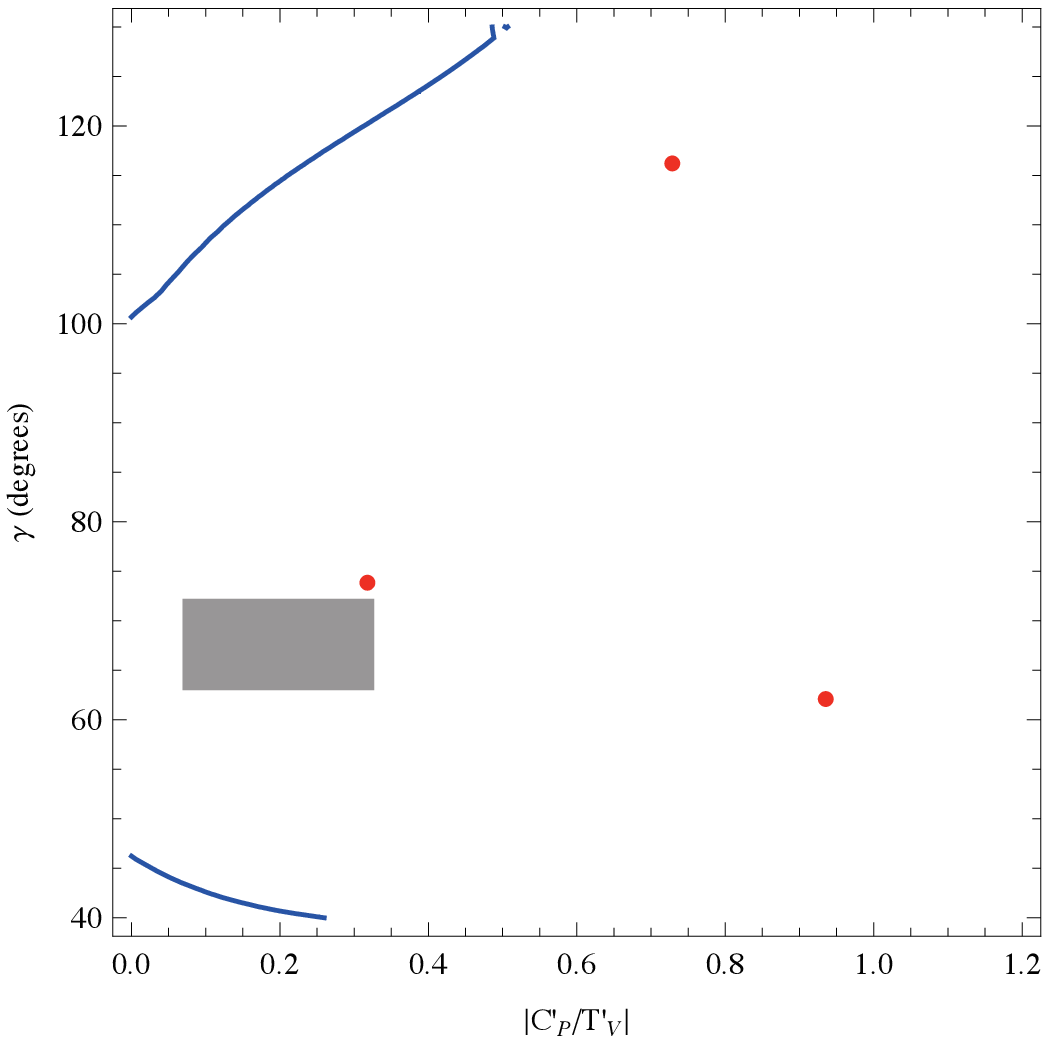}
\hspace{0.5cm}
\includegraphics[width=0.45\textwidth]{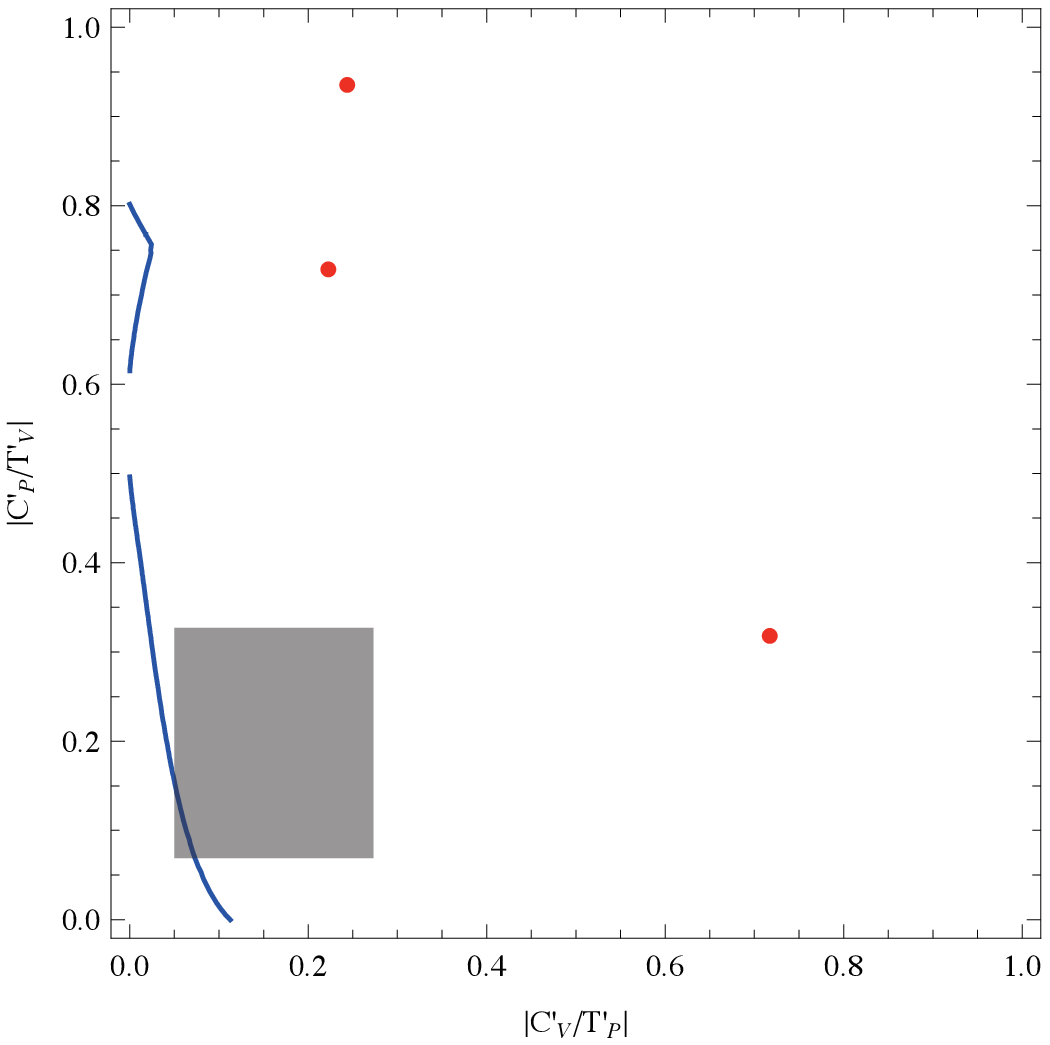}}
\caption{Fit2-a: correlation between the weak phase $\gamma$ and the ratio
  $|C'_P / T'_V|$ (left plot) or the ratios $|C'_P / T'_V|$ and $|C'_V / T'_P|$
  (right plot). The (red) dots correspond to the solutions listed in
  Table~\ref{tab:2-a}.  The contour of $\Delta\chi^2 = 1$ from the best solution
  ($\chi^2_{min} = 0.22$) is indicated by the solid (blue) curve.  The shaded
  region is given by the 1$\sigma$ ranges of Eqs.~(\ref{gammaval}) and
  (\ref{eq:QCDF}) for comparison.}
\label{fig-Fit2-a}
\end{figure}

The situation is even more striking in Fig.~\ref{fig-Fit2-a}, which shows the
correlation between the weak phase $\gamma$ and the ratio $|C'_P / T'_V|$ (left
plot) or the ratios $|C'_P / T'_V|$ and $|C'_V / T'_P|$ (right plot) for Fit2-a.
Taking the two plots together, one sees that all three central values ((red)
dots) are inconsistent with the shaded region, that is, they all indicate NP.
However, the shaded region is completely inside the ``solid region.'' Thus, the
inclusion of the fit errors removes all inconsistencies, so that all solutions
are in agreement with the SM.

\begin{figure}[thb]
\centering{%
\includegraphics[width=0.45\textwidth]{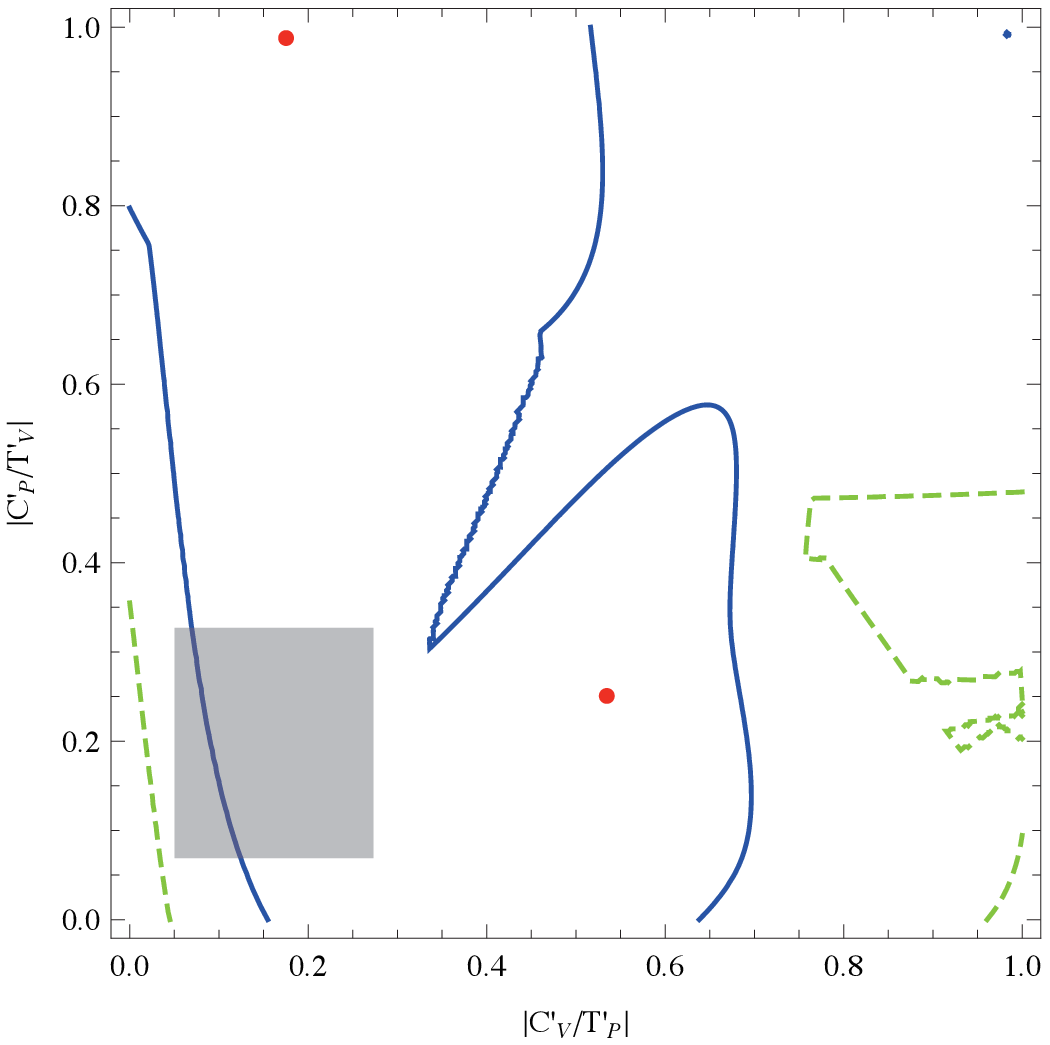}}
\caption{Fit2-b: correlation between the ratios $|C'_P /
  T'_V|$ and $|C'_V / T'_P|$. The (red) dots correspond to
  the solutions listed in Table~\ref{tab:2-b}.  The contours
  of $\Delta\chi^2 = 1$ and $2$ from the best solution
  ($\chi^2_{min} = 0.26$) are indicated by the solid (blue)
  and dashed (green) curves.  The shaded region is given by
  the 1$\sigma$ ranges of Eqs.~(\ref{gammaval}) and
  (\ref{eq:QCDF}) for comparison.}
\label{fig-Fit2-b}
\end{figure}

Finally, Fig.~\ref{fig-Fit2-b} shows the correlation between the ratios $|C'_P /
T'_V|$ and $|C'_V / T'_P|$ for Fit2-b.  Despite the fact that the central values
show some discrepancies with the shaded region, the inclusion of the fit errors
removes them, so that all solutions are in agreement with the SM.

{}From the figures, one sees that, when one includes the errors from the fits on
the theoretical parameters, all solutions are consistent with the SM, even if
the central values are not.  The reason for this is that the fit errors are
quite large.  And this is related to the fact that the experimental errors on
the $B \to K^* \pi$ and $B \to \rho K$ observables are also large, especially
those related to the CP asymmetries.  These errors are considerably larger than
those in $B \to K \pi$.  At present, it is not possible to clearly see if NP is
present in these $\btos$ $VP$ decays.  Hopefully, in the future, it will be
possible to increase the precision of the $B \to K^* \pi$ and $B \to \rho K$
measurements, so that we can further test the $\pi K$ puzzle.

In summary, we have performed fits to $B \to K^* \pi$ and $B \to \rho K$ decays
within the standard model (SM).  In particular, we have employed the relations
between electroweak-penguin and tree amplitudes to couple together the two types
of decay modes, which are otherwise separate.  Our fits are of two types: those
in which the weak phase $\gamma$ is extracted from the $B$-decay data, and those
in which an independent determination of $\gamma$ [Eq.~(\ref{gammaval})] is
imposed.  We begin our analysis by neglecting $P'_{uc}$.  We find that the
central value of $\gamma$ tends to be larger when no independent determination
is imposed (Fit1-a).  When Eq.~(\ref{gammaval}) is taken into account, two of
the solutions are consistent with the SM expectations, while the other has a
somewhat large $|C'_V/T'_P|$ ratio (Fit1-b).  Next we include $P'_{uc}$ in the
fits.  Without using the external constraint on $\gamma$, all three solutions
have a problem with too-large $|C' / T'|$ ratios.  In addition, one of these
gives a too-large value for $\gamma$ (Fit2-a).  They could indicate the presence
of new physics (NP).  After imposing Eq.~(\ref{gammaval}), there are two
solutions. One of these still has the problem of a large $|C'/ T'|$, while the
other is improved (Fit2-b).

These problems with the theory parameters potentially show some hints of NP.
However, the fits also yield large errors on these parameters. There is no
serious discrepancy with the predictions of the SM when these errors are taken
into account.  In order to make more decisive conclusions about NP in $B \to K^*
\pi$ and $B \to \rho K$, the uncertainties on the measurements of observables
have to be reduced, particularly for the CP asymmetries.  Such efforts will be
paid off by shedding additional light on the $B \to \pi K$ puzzle.

\bigskip

\noindent {\bf Acknowledgments}: C.C. would like to thank the hospitality of
NCTS-Hsinchu.  This work was financially supported by Grant
No.~NSC~97-2112-M-008-002-MY3 of Taiwan (CC) and by NSERC of Canada (DL).


\end{document}